\documentclass[%
 preprint,
superscriptaddress,
 amsmath,amssymb,
 aps,
 longbibliography
]{revtex4-2}

\usepackage{graphicx}
\usepackage{dcolumn}
\usepackage{bm}
\usepackage{amsfonts}
\usepackage{amssymb}
\usepackage{amsthm,amsmath}
\usepackage{mathrsfs}
\usepackage{indentfirst}
\usepackage{multirow}
\usepackage{subfigure}
\usepackage{upgreek}
\usepackage{hyperref}
\hypersetup{hypertex=true,
            colorlinks=true,
            linkcolor=blue,
            anchorcolor=blue,
            citecolor=blue}
\UseRawInputEncoding
\begin{document}

\title{Active control of particle position by boundary slip in inertial microfluidics}

\author{Chengliang Xuan}
\author{Weiyin Liang}
\author{Bing He}
\email{hebing@gxnu.edu.cn}
\author{Binghai Wen}
\email{oceanwen@gxnu.edu.cn}
\affiliation{%
 Guangxi Key Lab of Multi-Source Information Mining \& Security, Guangxi Normal University, Guilin 541004, China
}%
\affiliation{%
 School of Computer Science and Engineering, Guangxi Normal University, Guilin 541004, China
}%

\date{\today}
\begin{abstract}
Inertial microfluidic is able to focus and separate particles in microchannels based on the characteristic geometry and intrinsic hydrodynamic effect. Yet, the vertical position of suspended particles in the microchannel cannot be manipulated in real time. In this study, we utilize the boundary slip effect to regulate the parabolic velocity distribution of fluid in the microchannel and present a scheme to active control the vertical position of particles in inertial microfluidics. The flow field of a microchannel with a unilateral slip boundary is equivalent to that of the microchannel widened by the relevant slip length, and the particle equilibrium positions in the two microchannels are consistent consequently. Then, we simulate the lateral migrations of three kinds of typical particles, namely circle, ellipse, and rectangle in the microchannel. Unlike the smooth trajectories of circular particles, the motions of the elliptical and rectangular particles are accompanied by regular fluctuations and non-uniform rotations due to their non-circular geometries. The results demonstrate that the unilateral slip boundary can effectively control the vertical equilibrium position of particles. Thus, the present scheme enables to active manipulate the particles positions in vertical direction and can promote more accurate focusing, separating, and transport in inertial microfluidics.
\end{abstract}

\maketitle

\section{\label{sec1}INTRODUCTION}
Microfluidic is the technology of fluid manipulation in channels with dimensions of ten to hundreds of micrometers. It has emerged in recent years as a distinct new area of research thanks to its application in many diverse fields, such as chemistry, biology, medicine, and physics\cite{RN4236,RN4218,RN4312,RN4266,RN4265,RN4264,RN4268}. Nowadays, several technologies have already been proposed and developed to manipulate particles in microfluidic systems. According to the source of the manipulating forces, these technologies can be categorized as active and passive types. Active technologies rely on external force fields, whereas passive technologies depend entirely on the channel geometry or intrinsic hydrodynamic forces. As a passive technology, inertial microfluidic is used for particle focusing\cite{RN4304}, sorting\cite{RN4303}, and enriching\cite{RN4305} in industry, biology, and medicine. Inertial microfluidic originates from a wonderful natural phenomenon. As early as the 1960s, Segr$\rm{\acute{e}}$ and Silberberg found that suspended spherical particles in a pipe flow would migrate laterally away from the wall and finally reach a certain lateral equilibrium position\cite{RN4230}. The phenomenon was called the Segr$\rm{\acute{e}}$--Silberberg effect later and can also be observed on other shaped particles, such as cylinder, ellipse, disk, rod, biconcave particle, etc. Until 2007, Di Carlo \emph{et al.}\cite{RN4235} performed the inertial focusing of particles at a micro--scale. In a pipe with a diameter of the order of microns, the length of pipe required for particles to reach the equilibrium state was reduced to centimeters or even millimeters. The inertial migration of particles has now been practically applied in microfluidic.

Different focusing and separation effects were achieved through microchannels with different structures. The microchannel structure is the important parameter that determines the functionality and performance of inertial microfluidic devices. Ramachandraiah \emph{et al.}\cite{RN4248} used U--shaped and S--shaped microchannels to achieve the focusing of particles and found that the focusing position of particles is independent of the radius of curvature. Bhagat \emph{et al.}\cite{RN4284} used a 10--loop spiral microchannel to achieve 3--D focusing of fluorescently labeled 6 $\upmu$m particles without any additional sheath fluids. Sun \emph{et al.}\cite{RN3808} proposed a passive double spiral microfluidic device to continuously and efficiently separate and enrich tumor cells from diluted whole blood. Zhang \emph{et al.}\cite{RN4251} carefully evaluated the effect of particle centrifugal force on particle focusing in a serpentine microchannel and demonstrated for the first time that a single focusing streak can be achieved in a symmetric serpentine channel.

At the same time, when the characteristic length of the flow field goes to micron/nanometer, the interaction between the fluid and the wall must be considered. As a result, the boundary slip will occur at the wall and it has an important effect on the fluid field at micro--nano scale\cite{RN4311}. To quantify the boundary slip, Navier\cite{RN4302} proposed the concept of  `slip length', which is defined as the distance inside the wall at which the extrapolated fluid velocity would be equal to the velocity of the wall\cite{RN4293}. Several studies about the effect of the boundary slip have also been reported\cite{RN4297,RN4298}. Li \emph{et al.}\cite{RN4297} studied the effect of nanobubbles on the slippage experimentally and theoretically. In 5 $\upmu$m $ \times $ 5 $\upmu$m area, they found an increase from 8 to 512 nm in slip length by increasing the surface coverage of nanobubbles from 1.7$\%$ to 50.8$\%$ and decreasing the contact angle of nanobubbles from ${42.8^ \circ }$ to ${16.6^ \circ }$. Their results indicate that nanobubbles can always act as a lubricant and significantly increase the slip length. Minakov \emph{et al.}\cite{RN4298} studied the flow regimes and mixing performance in a T--type micromixer at high Reynolds numbers. They found that the flow regimes and the efficiency of mixing can be changed through using the slip boundary conditions.

Many researchers implemented a lot of numerical simulations on inertial microfluidic. In particular, the lattice Boltzmann method (LBM), as a reliable computational fluid dynamics method, is widely used to numerical simulations of microfluidic\cite{RN4290,RN3935,RN1617,RN4129,RN3116,RN4286,RN3368}. Sun \emph{et al.}\cite{RN3935} studied the particle focusing in a three--dimensional rectangular channel with the lattice Boltzmann method. Huang \emph{et al.}\cite{RN1617} used a multi--relaxation--time lattice Boltzmann method to study the rotation of a spherical particle in a Couette flow. They found seven periodic and steady rotation modes for a prolate spheroid. Wen \emph{et al.}\cite{RN4129,RN3116} simulated the migration of biconcave particles in straight channels and made a positive contribution to the study of blood circulation of birds with elliptical red blood cells. Liu \emph{et al.}\cite{RN4286} carried out a three-dimensional numerical simulation of the movement of particles in serpentine microchannel and proposed a fitting formula for the inertial lift on a sphere drawn from DNS data obtained in straight channels. However, there is no method to control particles' equilibrium positions in real time in inertial microfluidic. Study shows that slip exists at the microscale and its value is significantly affected by surface and fluid characteristics\cite{RN4309}. 

In this paper, the slip boundary condition is used on the side of Poiseuille flow. The velocity distribution of the flow field is controlled by adjusting the slip length, so as to control the equilibrium positions of particles. The simulation results show that this method can control the target particles more accurately and the equilibrium positions can be adjusted in real time. In addition, this method is very simple and robust. The structure of this paper is as follows. In section \ref{sec2}, we briefly describe the lattice Boltzmann method and moving boundary conditions. Section \ref{sec3} is devoted to describing the slip boundary condition, derivation, and verification of the slip length formula. The effect of boundary slip on the migration trajectory of different particles is presented in detail in Section \ref{sec4}. Section \ref{sec5} concludes the paper.

\section{\label{sec2}NUMERICAL METHOD}
To analyze the effect of the boundary slip, the lattice Boltzmann method with single relaxation time is used for simulating Poiseuille flow. And to describe precisely the motion of particles with different shape, the moving particle boundary is treated by the quadratic interpolation and the hydrodynamic force is calculated by the Galileo invariant momentum exchange method.

\subsection{\label{sec2.1}Lattice Boltzmann method}
Nowadays, the lattice Boltzmann method has developed into an alternative and promising numerical scheme for simulating complex fluid flows\cite{RN4318,RN3684}. Compared with other traditional numerical methods, this method combines the advantages of the macroscopic model and the molecular dynamics model. It has the advantages of simple description of fluid interaction, easy setting of complex boundaries, easy parallel computing, and easy implementation of program.

The lattice Boltzmann model with single relaxation time (SRT) can be written as\cite{RN4289,RN4290,RN3397}
\begin{eqnarray}\label{Eq. 1}
\frac{{\partial {f_i}}}{{\partial t}} + {{\bm{e}}_i} \cdot \nabla {f_i} =  - \frac{1}{\tau }\left( {{f_i} - f_i^{\left( {eq} \right)}} \right){\kern 1pt} {\kern 1pt} {\kern 1pt} {\kern 1pt} {\kern 1pt} {\kern 1pt} {\rm{     }}{\kern 1pt} \left( {i = 0,1,2, \ldots ,N - 1} \right){\rm{ }},
\end{eqnarray}
where ${{\bm{e}}_i}$ is a discrete velocity vector, ${f_i}$ is the particle distribution function with the velocity ${{\bm{e}}_i}$, $f_i^{\left( {eq} \right)}$ is the corresponding equilibrium distribution function, $N$ is the number of the different velocities in the model, and $\tau$ is the relaxation time. Eq. (\ref{Eq. 1}) is discretized in space ${\bm{x}}$ and time $t$:
\begin{eqnarray}\label{Eq. 2}
{f_i}({\bm{x}} + {{\bm{e}}_i}\delta t,t + \delta t) - {f_i}({\bm{x}},t) =  - \frac{1}{\tau }[{f_i}({\bm{x}},t) - f_i^{(eq)}({\bm{x}},t)]{\rm{ }},
\end{eqnarray}
where ${\delta _t}$ is the time step. In the model on a square lattice in two dimensions (D2Q9), the discrete velocity set is ${\bm{e}} = \{ (0, 0), (1, 0), (0, 1), ( - 1, 0), (0,  - 1), (1, 1), ( - 1, 1), ( - 1,  - 1), (1,  - 1)\} $ given by in nine directions respectively. $f_i^{(eq)}\left( {{\bm{x}},t} \right)$ can be calculated by
\begin{eqnarray}\label{Eq. 3}
f_i^{(eq)}({\bm{x}},t) = \rho {\omega _i}[1 + 3({{\bm{e}}_i} \cdot {\bm{u}}) + \frac{9}{2}{({{\bm{e}}_i} \cdot {\bm{u}})^2} - \frac{3}{2}{{\bm{u}}^2}]{\rm{ }},
\end{eqnarray}
where weight factors $\omega _i$ are given by ${\omega _o} = 4/9$, ${\omega _{1 - 4}} = 1/9$, ${\omega _{5 - 8}} = 1/36$, $\rho $ and ${\bm{u}}$ are the macroscopic density and the macroscopic velocity vector respectively, are given by
\begin{eqnarray}\label{Eq. 4}
\rho  = \sum\limits_i {{f_i}} \quad {\rm{and}} \quad {\bm{u}} = \frac{1}{\rho }\sum\limits_i {{{\bm{e}}_i}{f_i}}~.
\end{eqnarray}

The lattice Boltzmann method applies two essential steps, collision and streaming, to reveal phenomena at the mesoscopic scale. During a time step, the particle distribution functions in a lattice site collide and then flow into its neighboring lattice sites\cite{RN2089}. Hence, the corresponding computations of Eq. (\ref{Eq. 2}) are performed as
\begin{eqnarray}\label{Eq. 5}
&&{\rm{Collision:\quad}}{\tilde f_i}\left( {{\bm{x}},t} \right) - {f_i}\left( {{\bm{x}},t} \right) =  - \frac{1}{\tau }\left[ {{f_i}\left( {{\bm{x}},t} \right) - f_i^{\left( {eq} \right)}\left( {{\bm{x}},t} \right)} \right]{\rm{ }},
\\ \label{Eq. 6}
&&{\rm{Streaming:\quad}}{f_i}\left( {{\bm{x}} + {{\bm{e}}_i}\delta x,t + \delta t} \right) = {\tilde f_i}\left( {{\bm{x}},t} \right){\rm{ }},
\end{eqnarray}
where ${f_i}$ and ${\tilde f_i}$ denote precollision and postcollision states of the particle distribution functions, respectively. The dominant part of the computations, namely the collision step, is completely local, so the discrete equations are natural to parallelize. 
\subsection{\label{sec2.2}Moving boundary conditions}
The LBM has also been effectively applied to simulations of particulate suspensions in fluids. The curves boundaries of the particles are usually approximated by zig--zag staircase thus bounce--back boundary condition can be directly applied. Filippova and Hanel\cite{RN4004} proposed a curve boundary condition in 1998. Their method constructed a fictitious equilibrium distribution function for non--fluid nodes, so as to find the missing distribution function for fluid nodes on the boundary. Mei \emph{et al.}\cite{RN383} developed a second--order accurate treatment of the boundary condition for a curved boundary, which is an improvement of a scheme of Filippova and Hanel. Despite the success of these methods in the curved boundaries, there is no rigorous theory on the treatment of moving boundaries. Lallemand and Luo\cite{RN2521,RN602} proposed the quadratic interpolation method to treat the curved boundary and the moving particular boundary. In here, their method is adopted to treat with the moving boundary of particle in fluid.
\begin{figure}[!htbp]
\centering
\includegraphics[width=100mm]{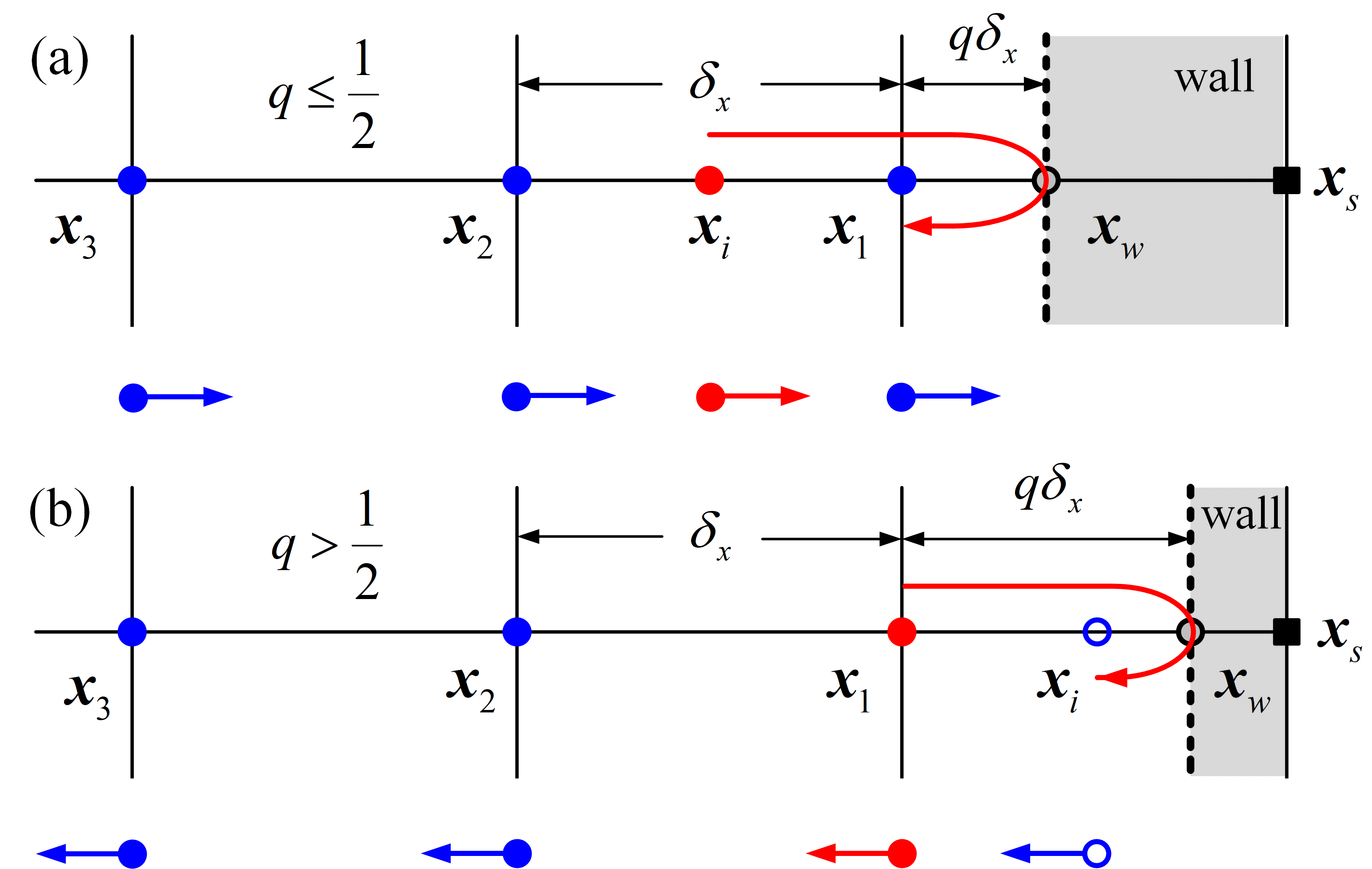}
\caption{Schematic diagram of curved fluid--solid boundary condition. (a) $q \le 1/2$ and (b) $q > 1/2$, where $q = \left( {{{\bm{x}}_1} - {{\bm{x}}_w}} \right)/\left( {{{\bm{x}}_1} - {{\bm{x}}_s}} \right)$.}
\label{Fig. 1}
\end{figure}

Schematic diagram of curved fluid-solid boundary condition is shown in FIG. \ref{Fig. 1}. The parameter $q$ defines the fraction in the fluid region of a grid spacing intersected by the boundary. The node ${{\bm{x}}_s}$ is the boundary node and ${{\bm{x}}_w}$ is the intersection point on the fluid-solid link. To avoid extrapolations, the scheme is divided into two parts according to the value of $q$. After collision and advection, the interpolation formulas are described as:
{\setlength\abovedisplayskip{1pt}
\setlength\belowdisplayskip{1pt}
\begin{small}
\begin{eqnarray}\label{Eq. 7}
\left\{ 
{\begin{array}{*{20}{l}}
{\!{f_{\bar i}}({{\bm{x}}_1},t)\! =\! q(1\! +\! 2q){{\tilde f}_i}({{\bm{x}}_1},t)\! +\! (1\! -\! 4{q^2}){{\tilde f}_i}({{\bm{x}}_2},t)\! -\! q(1\! -\! 2q){{\tilde f}_i}({{\bm{x}}_3},t)\! +\! 3{\omega _i}({{\bm{e}}_i} \cdot {{\bm{u}}_w}) \  {\rm{        }}q \le 1/2}\\
{\!{f_{\bar i}}({{\bm{x}}_1},t)\! =\! \frac{1}{{q(1 + 2q)}}{{\tilde f}_i}({{\bm{x}}_1},t) + \frac{{2q - 1}}{q}{f_{\bar i}}({{\bm{x}}_2},t) - \frac{{2q - 1}}{{2q + 1}}{f_{\bar i}}({{\bm{x}}_3},t) + \frac{{3{\omega _i}}}{{q(2q + 1)}}({{\bm{e}}_i} \cdot {{\bm{u}}_w})\quad\ \ {\rm{   }}q > 1/2}
\end{array}} 
\right.,
\end{eqnarray}
\end{small}
}
where ${\tilde f_i}\left( {{\bm{x}},t} \right)$ is the distribution function streamed from ${\bm{x}}$ in $i$ direction. ${{\bm{x}}_2}$ and ${{\bm{x}}_3}$ are two points adjacent to ${{\bm{x}}_1}$ along $i$ direction for interpolation, ${{\bm{u}}_w}$ represents the velocity of the moving boundary at the point of the intersection ${{\bm{x}}_w}$, and ${\omega _i}$ takes 2/9 for $i = 1, 2, 3, 4$ and 1/18 for $i = 5, 6, 7, 8$\cite{RN3873}. When $q \le 1/2$, interpolation calculation is performed before flow and rebound. When $q > 1/2$, interpolation calculation is performed after flow and rebound. 
\subsection{\label{sec2.3}Hydrodynamic force evaluation}
The hydrodynamic force in the lattice Boltzmann method can be efficiently evaluated by using a momentum exchange method. Ladd \emph{et al.}\cite{RN18} proposed the original momentum exchange method, which lay the particle boundary discretely and approximately at the middle of the link between a solid node and a fluid node, namely a fluid--solid link. A momentum item based on the boundary velocity was added to the distribution functions which were bounced back from the particle boundary, and the momentum--exchange occurred during the streaming step. Aidun \emph{et al.}\cite{RN25} improved Ladd's model by directly representing the solid particle without fluid inside. The momenta of the covered and uncovered nodes were involved in the force evaluation for moving solid particles. However, their effect was not investigated in detail. A common drawback in Ladd's and Aidun's method is that the boundary geometry, which is located at the middle of fluid--solid links, is zigzag. Mei \emph{et al.}\cite{RN23} introduced the curved boundary conditions in the momentum--exchange method so that the particulate geometry could be accurately represented on the grid level.
\begin{figure}[!htbp]
\centering
\includegraphics[width=100mm]{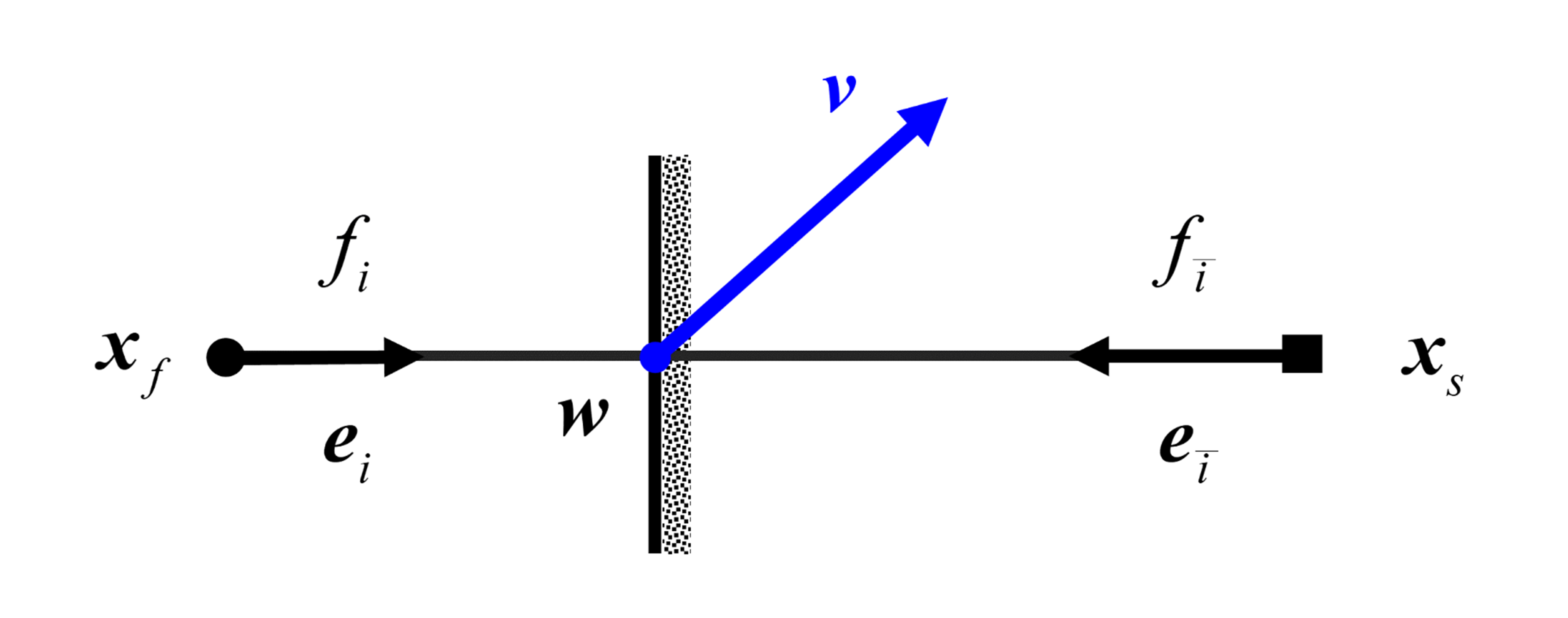}
\caption{Schematic diagram to illustrate a moving boundary crossing a fluid--solid link at the point of intersection ${\bm{w}}$. ${{\bm{x}}_f}$ and ${{\bm{x}}_s}$ denote the adjacent fluid and boundary nodes. The boundary has a velocity ${\bm{v}}$ at the point ${\bm{w}}$.}
\label{Fig. 2}
\end{figure}

FIG. \ref{Fig. 2} presents a moving boundary is located between a fluid node ${{\bm{x}}_f}$ and a boundary node ${{\bm{x}}_s}$. The boundary has a vector velocity ${\bm{v}}$ at the point of intersection ${\bm{w}}$. Wen \emph{et al.}\cite{RN3268,RN3482} introduced the relative velocity into the interfacial momentum transfer to compute the hydrodynamic force and proposed a Galilean invariant momentum (GME) exchange equation:
\begin{eqnarray}\label{Eq. 8}
{\bm{F}}({{\bm{x}}_s}) = ({{\bm{e}}_i} - {\bm{v}}){f_i}({{\bm{x}}_f},t) - ({{\bm{e}}_{\bar i}} - {\bm{v}}){f_{\bar i}}({{\bm{x}}_s},t)~.
\end{eqnarray}

It is demonstrated to greatly enhance the computational accuracy and robustness of moving boundaries in the dynamic fluid. Especially, the algorithm meets full Galilean invariance and is independent of boundary geometries. The total hydrodynamic force and torque are calculated by
\begin{eqnarray}\label{Eq. 9}
{\bm{F}} = \sum {{\bm{F}}({{\bm{x}}_w})}
\end{eqnarray}
and
\begin{eqnarray}\label{Eq. 10}
{\bm{T}} = \sum {({{\bm{x}}_w} - {\bm{R}}) \times {\bm{F}}({{\bm{x}}_w})} ,
\end{eqnarray}
where ${\bm{F}}$ and ${\bm{T}}$ are the summation of force and torque on each ${{\bm{x}}_w}$, ${\bm{R}}$ is the mass center of the solid particle. GME is simple, efficient in computation, and clearly expressed physically\cite{RN3268}.
\section{\label{sec3}Regulating the velocity distribution by boundary slip}
The nature of the boundary condition for fluid flows past solid surfaces has been a subject of interest for a long time. No--slip boundary condition, that is, the velocity of a liquid at a surface is always identical to the velocity of the surface, is extremely successful in describing macro--scale viscous flows. However, at nanoscale, this assumption is usually broken down\cite{RN4312}. Some early experiments that indicating slip mostly involving the flow of liquids through thin lyophobic capillaries\cite{RN4314,RN4313}. Some new experiments using more modern technology have also shown evidence of boundary slip\cite{RN4315,RN4316,RN4317}.
\subsection{\label{sec:3.1}Slip boundary scheme}
\begin{figure}[!htbp]
\centering
\includegraphics[width=150mm]{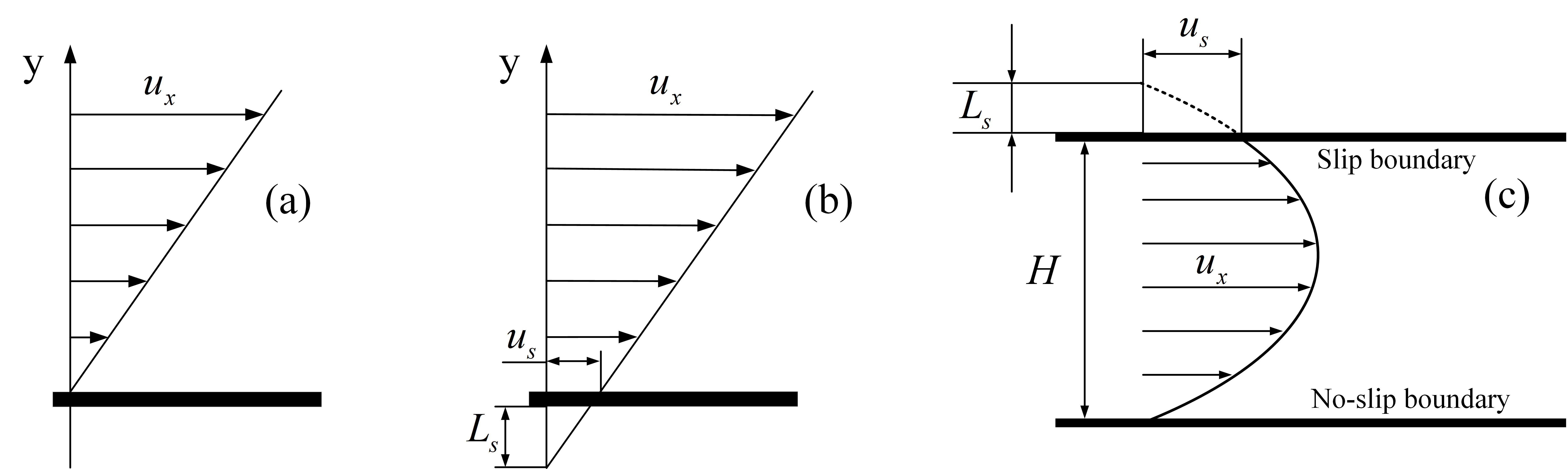}
\caption{Schematic diagrams of shear flows with (a) a no--slip boundary and (b) a slip boundary. (c) A channel flow with a slip upper boundary and a no--slip lower boundary. Owing to the boundary slip, the position of maximum velocity in the channel is moved towards the upper boundary, and the deformed parabolic velocity distribution is equivalent to that of the Poiseuille flow in a channel with width $H + {L_s}$.}
\label{Fig. 3}
\end{figure}

As early as 1823, Navier\cite{RN4302} proposed the linear slip boundary condition hypothesis, which assumes that the slip velocity is proportional to the local shear rate,
\begin{eqnarray}\label{Eq. 11}
{u_s} = {L_s}\frac{{\partial {u_x}}}{{\partial y}}\left| {_{wall}} \right.,
\end{eqnarray}
where ${u_s}$ is the slip velocity on the boundary, ${L_s}$ is the slip length and ${u_x}$ is the tangential velocity of fluid along the boundary surface. FIG. \ref{Fig. 3}(a) and (b) show the shear flow with a no--slip boundary and a slip boundary. When a channel flow with a slip upper boundary and a no--slip lower boundary, the velocity distribution of the flow field is shown in FIG. \ref{Fig. 3}(c). In this paper, we adopt the scheme of the FIG. \ref{Fig. 3}(c) to simulate migration of particles.

In numerical simulations, boundary slip is usually implemented through boundary conditions. Especially, a kind of kinetic boundary conditions has been developed in LBM to simulate boundary slip phenomena effectively. Succi\cite{RN4292} proposed a combination of the bounce--back and specular reflection condition to capture slip velocity on the solid wall, which is denoted as Bounce--back Specular Reflection method or BSR method. Guo \emph{et al.}\cite{RN4294} analyzed numerical error and discrete effect on the bounce--back and specular--reflection boundary condition and the Maxwellian boundary condition, and found that both schemes are virtually equivalent in principle. Chai \emph{et al.}\cite{RN4308} proposed a new combination of bounce--back and full diffusive boundary condition to investigate the incompressible gaseous flow in a microchannel with surface roughness. In this paper, we adopt the hybrid boundary condition proposed by Succi, which is simple, efficient and has been widely applicated.
\begin{figure}[!htbp]
\centering
\includegraphics[width=100mm]{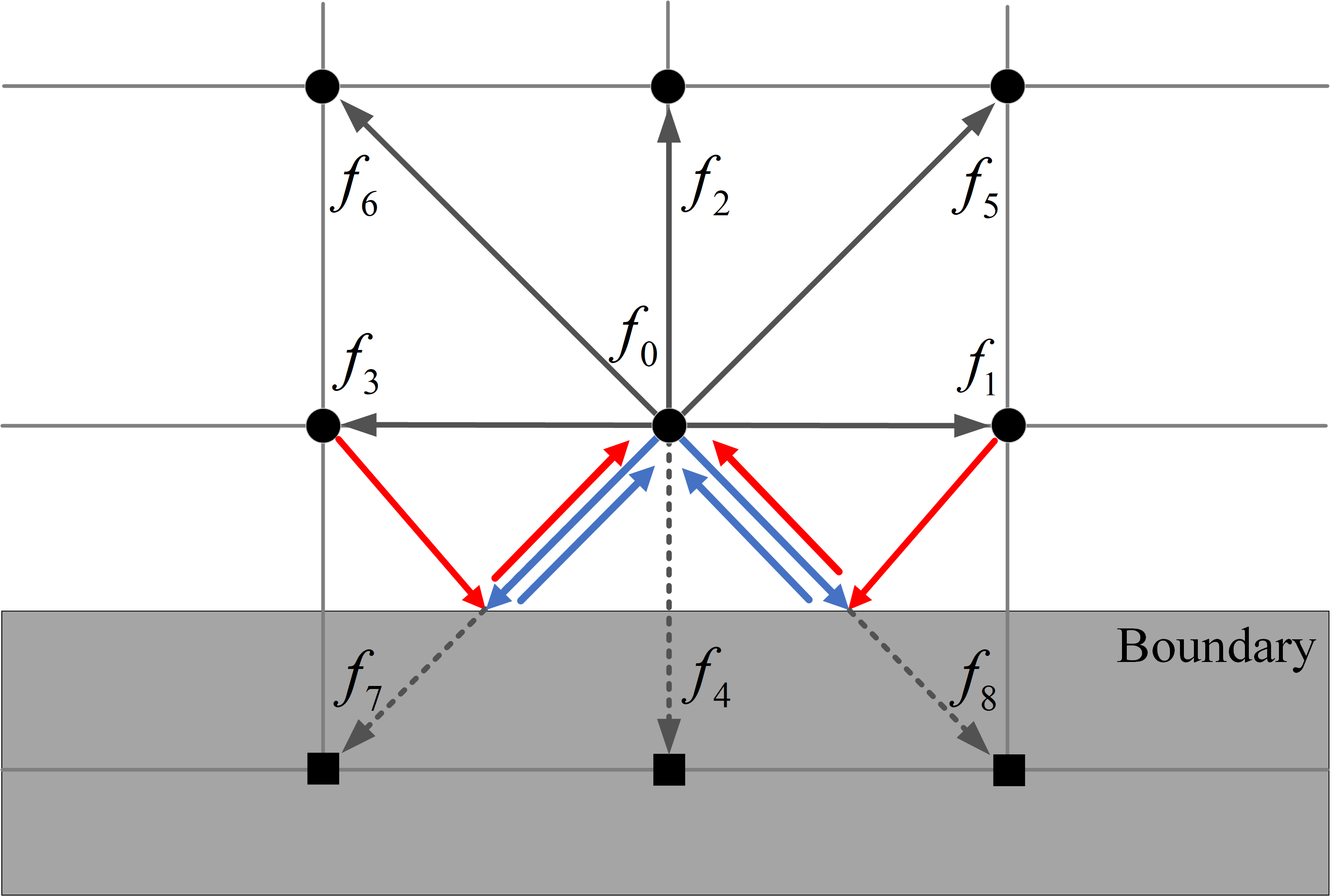}
\caption{Schematic diagram of bounce--back specular reflection boundary condition. The boundary is located at the middle of the two rows of lattices, and the distribution functions to be solved are composed of the bounce--back and the specular reflection in a proportional factor $r$. The blue lines represent bounce--back and the red lines represent specular reflection.}
\label{Fig. 4}
\end{figure}

As shown in FIG. \ref{Fig. 4}, the distribution functions of ${f_2}$, ${f_5}$, ${f_6}$ for the grid need to be obtained according to the boundary conditions, others can be obtained according to the migration step. The hybrid boundary condition proposed by Succi combines the bounce--back (the blue lines) with the specular reflection (the red lines) based on proportional factor $r$. For the BSR method, the unknown distribution functions are given by
\begin{eqnarray}\label{Eq. 12}
\left\{ {\begin{array}{*{20}{c}}
{{f_2} = {{f'}_4}}\\
{{f_5} = r{{f'}_7} + (1 - r){{f'}_8} + 2r\rho {\omega _i}{c_5} \cdot {{\bm{u}}_\omega }/2c_s^2}\\
{{f_6} = r{{f'}_8} + (1 - r){{f'}_7} + 2r\rho {\omega _i}{c_6} \cdot {{\bm{u}}_\omega }/2c_s^2}
\end{array}} \right.,
\end{eqnarray}
where ${f'_i}$ is the distribution function of the node in the $i$ direction after collision, ${{\bm{u}}_\omega }$ is the velocity of the wall, $r$ represents the proportion of bounce--back reflections in the interactions with the wall and $1 - r$ represents the proportion of specular reflections. Therefore, $r = 1$ corresponds to pure bounce--back reflection and $r = 0$ to pure specular reflection.
\subsection{\label{sec3.2}Slip length of Poiseuille flow}
For the incompressible Newtonian fluid with constant viscosity, one--dimensional steady Poiseuille flow is carried out in the x--direction between two infinite plates. If the continuity assumption is satisfied and the z--direction is unlimited width, the Navier--Stokes equation in the Cartesian coordinate system can be simplified as
\begin{eqnarray}\label{Eq. 13}
\frac{{{d^2}{u_x}}}{{d{y^2}}} = \frac{1}{\mu }\frac{{dp}}{{dx}},
\end{eqnarray}
where $\mu$ is the viscosity of the fluid and $p$ is the pressure.

As shown in FIG. \ref{Fig. 3}(c), the lower boundary is no--slip ${u_{x\left| {y = 0} \right.}} = 0$, whereas the upper boundary has the slip velocity ${u_{x\left| {y = H} \right.}} = {u_s}$. Substituting the two boundary conditions into Eq. (\ref{Eq. 13}) gives the velocity distribution in the vertical direction
\begin{eqnarray}\label{Eq. 14}
{u_x}\left( y \right) = \frac{y}{H}{u_s} + \frac{1}{{2\mu }}\frac{{dp}}{{dx}}({y^2} - Hy){\rm{ ,}}
\end{eqnarray}
where $y \in \left( {0,H} \right)$. The analytical solution of Poiseuille flow with flow field width $H + {L_s}$ is:
\begin{eqnarray}\label{Eq. 15}
{u_x}\left( y \right) = \frac{1}{{2\mu }}\frac{{dp}}{{dx}}({y^2} - Hy - {L_s}y)~.
\end{eqnarray}

Based on Eq. (\ref{Eq. 14}) and the analytical solution Eq. (\ref{Eq. 15}), we can obtain the relationship between the slip length ${L_s}$ and the slip velocity ${u_s}$:
\begin{eqnarray}\label{Eq. 16}
{L_s} =  - \frac{{2\mu {u_s}}}{H}{(\frac{{dp}}{{dx}})^{ - 1}}~.
\end{eqnarray}

Then, according to the definition of velocity $\rho {u_j} = c\left( {f_1^j - f_3^j + f_5^j - f_6^j + f_8^j - f_7^j} \right) + \frac{{{\delta _t}}}{2}\rho a$ and the stream law of particle distribution function, the relation formula of the flow velocity in the x--direction between the adjacent mesh ${u_H}$ and ${u_{H - 1}}$ can be obtained\cite{RN4294}:
\begin{eqnarray}\label{Eq. 17}
{u_{H - 1}} = \frac{{1 - 2\tau  + 2r(\tau  - 2)}}{{1 - 2\tau  + 2r(\tau  - 1)}}{u_H} + \frac{{6(2\tau  - 1) + r(8{\tau ^2} - 20\tau  + 11)}}{{(2\tau  - 1)[1 - 2\tau  + 2r(\tau  - 1)]}}{\bm{a}}{\rm{ }},
\end{eqnarray}
where ${\bm{a}}$ is the external force. Consider the linear velocity distribution of the flow field, the external force ${\bm{a}}$ is 0. Eq. (\ref{Eq. 17}) can be simplified as:
\begin{eqnarray}\label{Eq. 18}
{u_{H - 1}} = \frac{{1 - 2\tau  + 2r(\tau  - 2)}}{{1 - 2\tau  + 2r(\tau  - 1)}}{u_H}~.
\end{eqnarray}

Substituting Eq. (\ref{Eq. 14}) and Eq. (\ref{Eq. 16}) into Eq. (\ref{Eq. 18}), we get the function of the slip length ${L_s}$, which is related to the rebound coefficient $r$.
\begin{eqnarray}\label{Eq. 19}
{L_s} =  - \frac{{(H - 1)[1 - 2\tau  + 2r(\tau  - 1)]}}{{(2\tau  - 1) + 2r(H - \tau  + 1)}}~.
\end{eqnarray}

\subsection{\label{sec3.3}Numerical verifications}
\begin{figure}[!htbp]
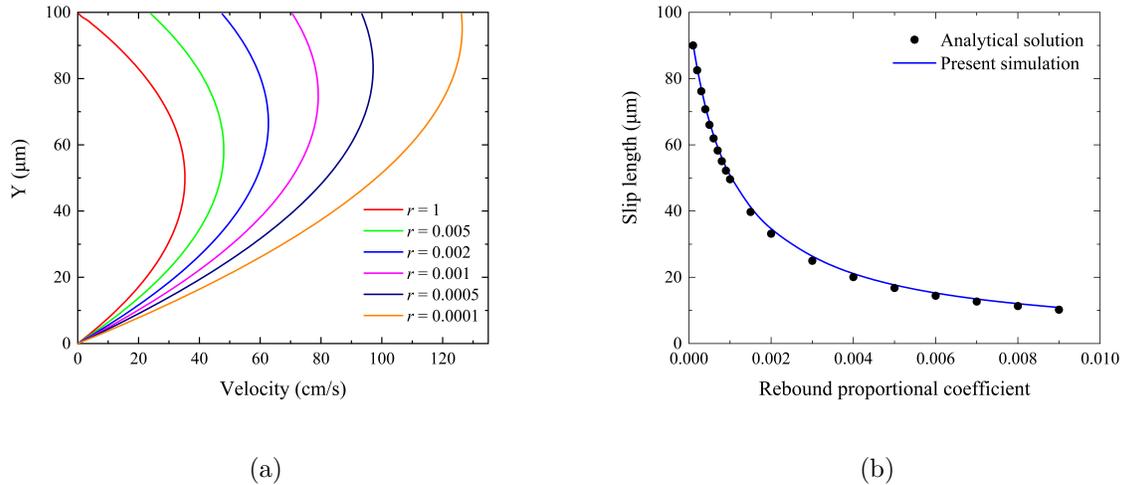

\centering
\subfigure[]{\includegraphics[width=8cm]{Fig_5a}}
\subfigure[]{\includegraphics[width=8cm]{Fig_5b}}
\caption{(a) The effect of boundary slip to velocity distribution of the channel flow at different rebound proportional coefficients $r$. (b) The slip lengths of the simulations with a series of rebound proportional coefficients are in agreement with the analytical solutions.}
\label{Fig. 5}
\end{figure}

In the above section, we introduced the slip boundary scheme and derived the relationship among the slip length, the slip velocity and rebound coefficient. In this subsection, the effectiveness of the BSR boundary condition in a horizontal channel is verified by the numerical experiment. The length ${l_0}$ and width ${h_0}$ of the channel are 1000 $\upmu$m and 100 $\upmu$m respectively. The fluid density is 1 ${\rm{g/c}}{{\rm{m}}^{\rm{3}}}$ and the kinematic viscosity coefficient is $\upsilon  = 1 \times {10^{ - 6}}\ {{\rm{m}}^2}{\rm{/s}}$. BSR boundary condition is applied at the upper boundary of the channel wall and half--way bounce--back boundary condition is applied at the lower boundary. As shown in FIG. \ref{Fig. 5}(a), the different velocity distribution of the flow field can be obtained by changing the coefficient $r$ in the slip boundary condition. When $r$ = 1, the slip boundary condition is equal to half--way bounce--back boundary condition, so the velocity of the flow field is symmetric. As the coefficient $r$ gradually decreases, the maximum velocity position of the flow field will gradually move towards the upper boundary, which is just like broadening the boundary of the flow field upward and the widened length is the slip length. The maximum velocity position of the flow field will only be infinitely close to the upper boundary but cannot exceed it. Also, the velocity of the flow field will become faster.

Then, in order to verify the correctness of the slip length formula in the previous section, the simulation with different rebound proportional coefficients is performed and the result is shown in FIG. \ref{Fig. 5}(b). The solid line is the slip length simulated by changing the coefficient $r$ of the BSR boundary condition. In this simulation, the slip length can reach 90 $\upmu$m when the coefficient $r$ is $1 \times {10^{ - 4}}$. Minakov \emph{et al.}\cite{RN4298} also obtains the similar size of slip length in their model. The result is consistent with that derived by the analytical solution and indicates the proposed slip length formula is correct. This also confirms that it is feasible to adjust the slip length by using slip boundary conditions.

\begin{figure}[!htbp]
\centering
\includegraphics[width=100mm]{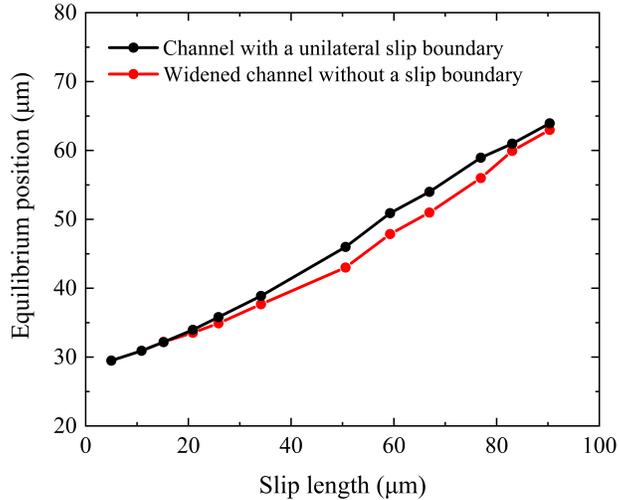}
\caption{The particle equilibrium positions in the channel with the width ${h_0}$ and a unilateral slip length ${l_s}$ agree well with those in the channel widened to ${h_0} + {l_s}$ without a slip boundary.}
\label{Fig. 6}
\end{figure}

To investigate the effect of boundary slip to the equilibrium position of the particle, a circular particle with a diameter of 20 lattice units, is added into the Poiseuille flow with slip boundary condition. Setting the boundary condition of the flow field, two cases are considered. The first one uses the slip boundary condition and sets different slip lengths ${l_s}$ on the upper boundary, and the half--way bounce condition is applied on the lower boundary. In the second case, half--way bounce condition is used in the upper and lower boundaries, the channel is widened to ${h_0} + {l_s}$. As shown in FIG. \ref{Fig. 6}, the black line is the simulation of the first case, and the red is the result of the second case. The results show that the boundary slip causes the maximum velocity point of the flow field to move upward, so that the equilibrium position of the particles moves upward. The final equilibrium positions of circle particles are basically consistent. When the slip length is less than 30 $\upmu$m, the equilibrium positions of two cases are fit perfectly. As the slip length grows from 40 to 80 $\upmu$m, the difference of the equilibrium positions between the two cases is about 5\%. When the slip length is large enough, the equilibrium position of the particle in the first case is relatively close to the upper wall. The main role of the wall is to slow down particles and keep them away from the wall, so the effect of the wall will be more obvious when the particle nearer to the wall. Thus, the difference decreases again after the slip length is larger than 80 $\upmu$m.
\section{\label{sec4}Active control of particle positions}
From the above results, it can be found that the velocity distribution of the flow field can be changed by employing the slip boundary conditions on the boundary of the Poiseuille flow. The change of velocity distribution in the flow field will affect the migration and the equilibrium position of the particle. According to Eq. (\ref{Eq. 19}), we can arbitrarily change the slip length by changing the coefficient $r$ of the slip boundary condition to control the equilibrium position of the particle. In this way, we present a scheme to active control the particle position in inertial microfluidics. The motion of particles can be controlled without the external force or the channel with special geometry.
\begin{figure}[!htbp]
\centering
\includegraphics[width=100mm]{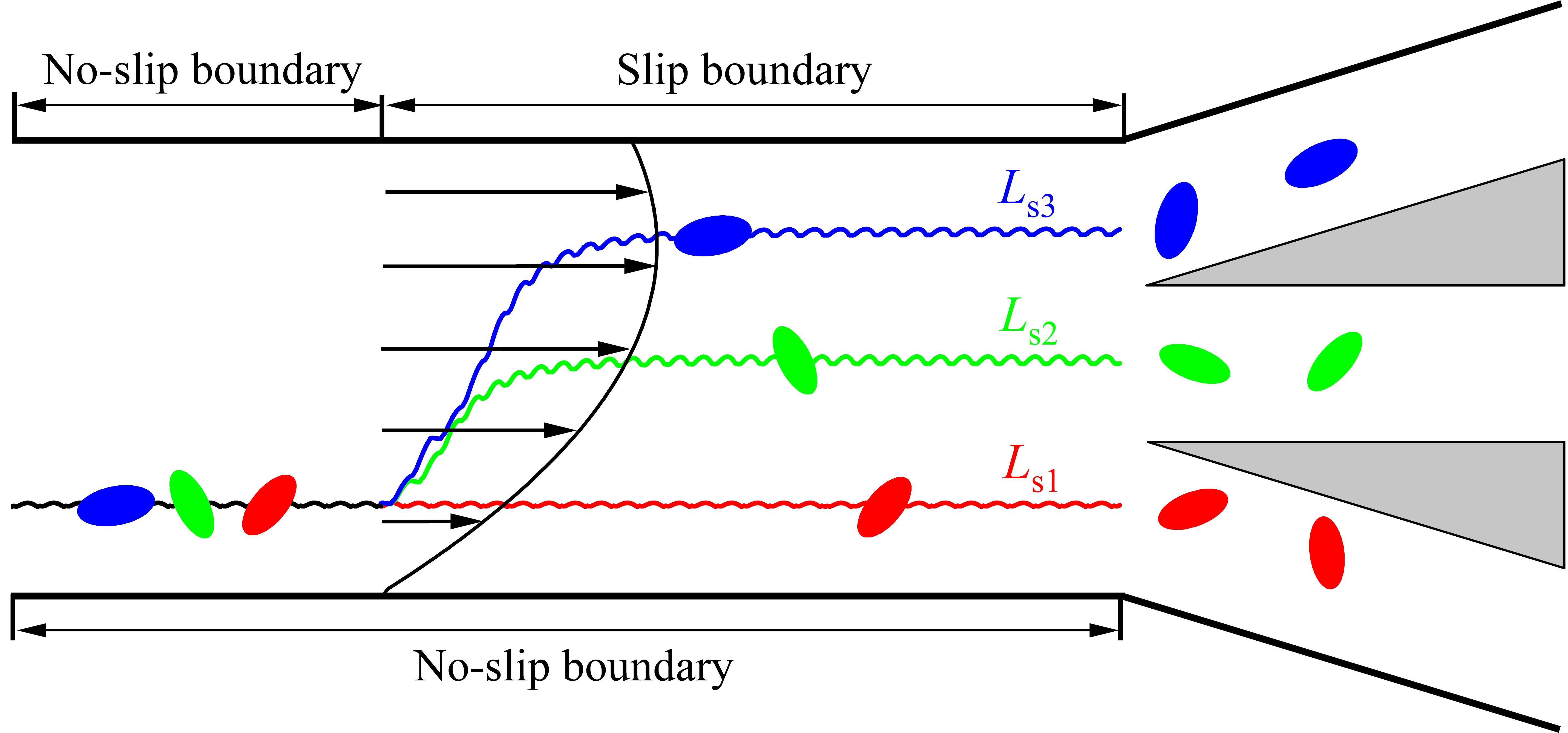}
\caption{Schematic diagram of active control of particle positions in inertial microfluidics. With the increasing slip length of the upper boundary, the red, green and blue particles can migrate to the different equilibrium positions in the vertical direction. The three slip lengths are ${L_{s1}}$, ${L_{s2}}$, ${L_{s3}}$ and satisfy ${L_{s1}} = 0$, ${L_{s1}} < {L_{s2}} < {L_{s3}}$.}
\label{Fig. 7}
\end{figure}

FIG. \ref{Fig. 7} depicts the scheme to active control of particle equilibrium positions by using the slip boundary condition at the upper boundary in Poiseuille flow. Our simulations are carried out in a two--dimensional rectangular domain 1000 $\times$ 100 (lattice units). The corresponding macroscopic width of the channel is 100 $\upmu$m. The fluid density is $1 \times {10^3}\ {\rm{ kg/}}{{\rm{m}}^{\rm{3}}}$ and the kinematic viscosity coefficient is $\upsilon  = 1 \times {10^{ - 6}}{\rm{ }}{{\rm{m}}^{\rm{2}}}{\rm{/s}}$, the density of particles is equal to the fluid. Reynolds number (Re) is a dimensionless number to characterize fluid flow and is expressed as ${\mathop{\rm Re}\nolimits}  = HU/\upsilon $, where $U$ is the mean fluid velocity in Poiseuille flow without a particle. The pressure boundary condition is applied at inlet and outlet of the channel. Half--way bounce--back boundary condition is applied at the first half of the upper boundary, so that the particles will focus to the same equilibrium position in the first half of the channel. The BSR slip boundary condition is implemented in the latter half of the upper boundary. Different slip lengths are set to controlling the equilibrium position of particles and ultimately forcing the particles migrate into different branches.

To make a detailed analysis of active controlling particle equilibrium position by boundary slip, particles with different shapes will be considered. 
\subsection{\label{sec4.1}Circular particle}
\begin{figure}[!htbp]
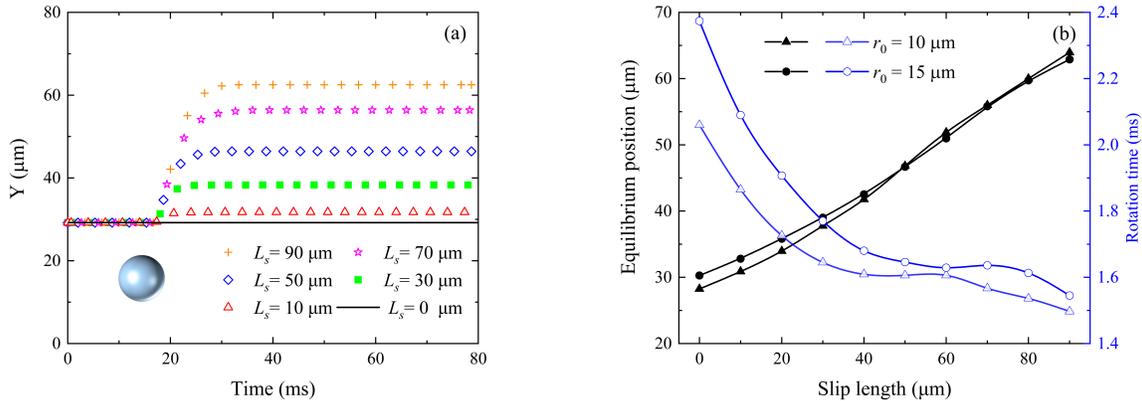

\centering
\subfigure{\includegraphics[width=8cm]{Fig_8a}}
\subfigure{\includegraphics[width=8cm]{Fig_8b}}
\caption{The migrations and equilibriums of circular particles in the channel flows with various unilateral slip lengths. (a) the migration trajectories and (b) the vertical equilibrium positions and the rotating periods of the circular particles, ${r_0}$ is the radius of the circular particle.}
\label{Fig. 8}
\end{figure}
Circular particles with the diameter of 20 and 30 $\upmu$m are added to this flow field. The migration trajectories, horizontal velocity, and period time of the circular particles when the slip length is 0, 10, 30, 50, 70, and 90 $\upmu$m are shown in Fig. \ref{Fig. 8}. The Reynolds number of Poiseuille flow is 24 when the slip length is 0 $\upmu$m. The migration trajectories of the circular particle with the diameter of 30 $\upmu$m are shown in FIG. \ref{Fig. 8}(a). The straight black line is the migration trajectory when the upper channel is no--slip, namely the classical Segr$\rm{\acute{e}}$--Silberberg effect. The process from placing the particles to reaching the equilibrium position in the first half of the channel is omitted. With the increase of the slip length, the velocity distribution of the flow field will change. The circular particles will gradually migrate upward and reach the equilibrium position.

FIG. \ref{Fig. 8}(b) shows the vertical equilibrium positions and rotating periods of circular particles changing with slip lengths. When the slip length increase from 0 to 90 $\upmu$m, the equilibrium positions of the circular particles increase linearly, gradually move towards the upper boundary. On the contrary, the period time becomes shorter. However, when the slip length is 50--70 $\upmu$m, the period time of the circular particle becomes basically the same. It is obvious that the larger particle, the longer the rotating period time is. The size of particle has little influence on the vertical equilibrium position.
\subsection{\label{sec4.2}Elliptical particle}
\begin{figure}[!htbp]
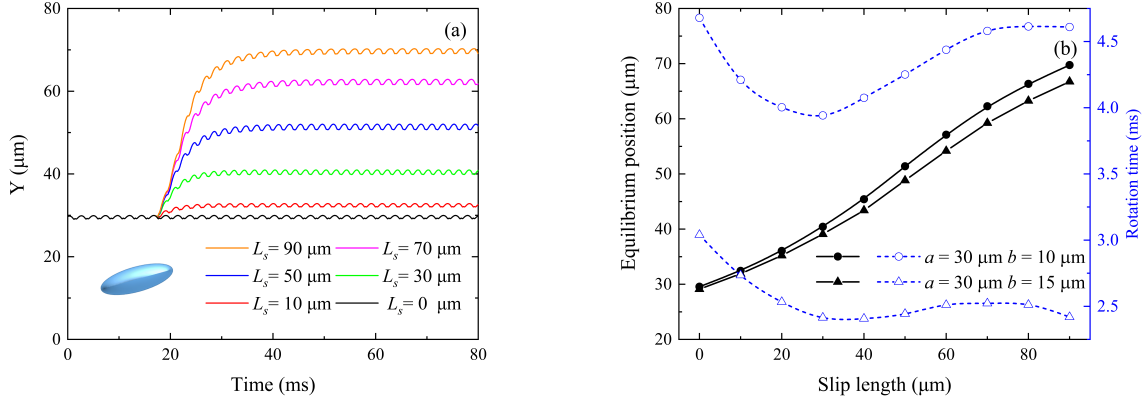

\centering
\subfigure{\includegraphics[width=8cm]{Fig_9a}}
\subfigure{\includegraphics[width=8cm]{Fig_9b}}
\caption{The migrations and equilibriums of elliptical particles in the channel flows with various unilateral slip lengths. (a) the migration trajectories and (b) the vertical equilibrium positions and the rotating periods of the elliptical particles, $a$ and $b$ are the long axis and short axis of the elliptical particle.}
\label{Fig. 9}
\end{figure}
The motion of two elliptical particle in different size is investigated under the same condition. The long axis of two elliptical particles is 30 $\upmu$m, and short axis of the elliptical particles are 10 and 15 $\upmu$m, respectively. As shown in FIG. \ref{Fig. 9}(a), due to the non--circular geometry of elliptical particles and the parabolic velocity distribution in the channel, the motions of the elliptical particles are accompanied by complex rotation and oscillations. Similar to circular particles, elliptical particles will reach different equilibrium positions with different slip lengths. In here, the equilibrium position is defined as the average position in a rotation period after the particle is in the equilibrium.

The vertical equilibrium positions and the rotating periods changing with the slip length of elliptical particles are shown in FIG. \ref{Fig. 9}(b). The influence of slip length on the vertical equilibrium positions and period time of elliptical particles is similar to that of circular particles. The influence of the size and aspect ratio of particles on equilibrium positions is not distinct. However, it is very clear that the particle with the larger aspect ratio has shorter rotation period and rotates faster. Because of the non--circular geometry, the area of thrust surface of the elliptical particle is uneven in the flow field. So, the same hydraulic force pushes the flatter elliptical particle to rotate more slowly. This is consistent to the observations in the related study\cite{RN4129}. The period times of the elliptical particle with different slip lengths all are greater than 2.4 ms and the maximum reaches 4.7 ms, while the maximum of that of the circular particle is only 2.4 ms even though its area is bigger than the ellipse.
\begin{figure}[!htbp]
\centering
\includegraphics[width=100mm]{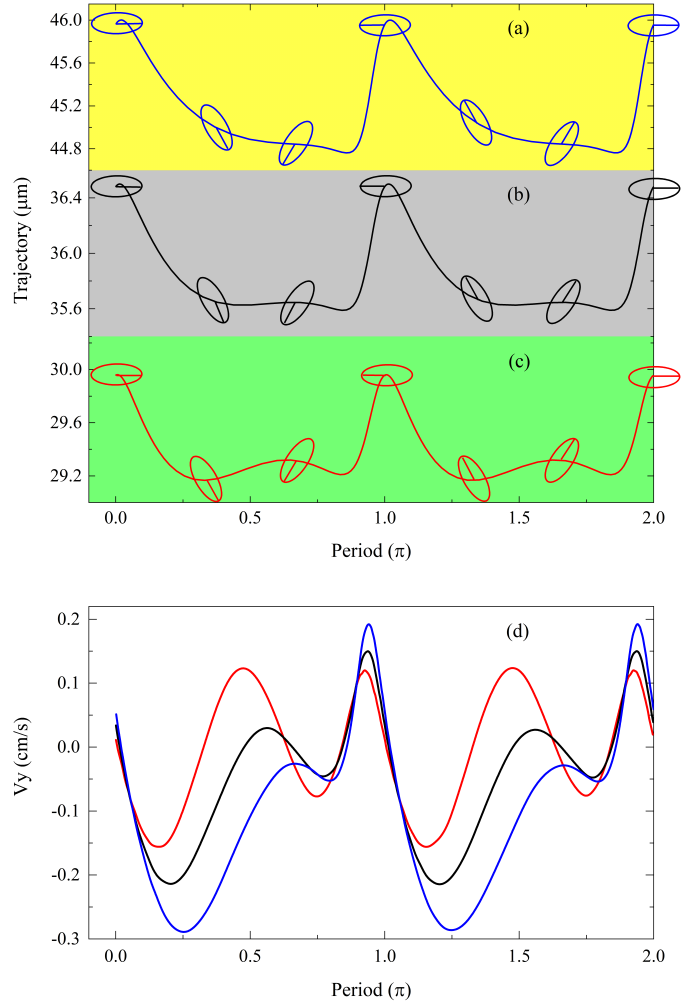}
\caption{(a) The migration trajectory and angular orientation of the elliptical particle with slip length of 40 $\upmu$m, (b) 20 $\upmu$m and (c) 0 $\upmu$m; (d) the linear velocity of the elliptical particles with slip length of 0 (red line), 20 (black line) and 40 (blue line) $\upmu$m.}
\label{Fig. 10}
\end{figure}

FIG. \ref{Fig. 10}(a)--(c) depict the trajectories and angular orientation of the elliptical particle with the slip length of 40, 20 and 0 $\upmu$m, respectively. In here, the clockwise is defined as the positive direction to investigate the rotation of the particle. The steering angles are 0, 1/3$\uppi$, 2/3$\uppi$, $\uppi$, 4/3$\uppi$, 5/3$\uppi$, and 2$\uppi$, respectively. The migration trajectory of the elliptical particle is like a saddle. The saddle shape can be considered as the combined action of the hydrodynamic force, the wall effect and the periodic oscillation. In steady state, the lateral migration amplitude of elliptic particles inside a rotation period is 0.8 $\upmu$m at ${L_s} = 0$ $\upmu$m. As the slip length increases to 40 $\upmu$m, the lateral migration amplitude increases to 1.2 $\upmu$m. It indicates that the lateral migration amplitude of elliptic particles inside a rotation period increases significantly with the increase of the slip length. When the slip length is 0 $\upmu$m, the fluctuation amplitude of the elliptic particle has a small wave peak during the steering angle 0.25 -- 0.75$\uppi$ in the half rotation period. However, the small wave peak gradually disappear as the slip length increases from 0 to 40 $\upmu$m. So, the oscillation trajectories of the particles will also change accordingly with the increase of the slip length.

FIG. \ref{Fig. 10}(d) shows vertical velocities of the elliptical particle with different slip lengths. The patterns of the linear velocity of ellipses with different slip lengths are similar within a single period. It suggests that the elliptical particles with the same aspect ratio have similar migration trajectories at different slip lengths. When the period is 0.25 -- 0.5$\uppi$ and 1.25 -- 1.75$\uppi$, the linear velocity gradually changes from negative to positive. However, the linear velocity remains negative in this interval when the slip length is greater than 0 $\upmu$m, because the boundary slip causes the velocity of flow field and the rotation speed of particle to increase so that the particle completes a quarter of the cycle before the linear velocity becomes positive. This confirms that the small wave peak disappear as the slip length increases gradually in FIG. \ref{Fig. 10}(a)--(b). Observing the linear velocity of the particle in the fluctuating process, it can be found that the elliptical particle migrates with a large cross--stream velocity in vertical direction if the slip length is large enough. These indicate that the oscillation amplitude of the elliptical particle with a long slip length is also more obvious than the shorter one. 
\subsection{\label{sec4.3}Rectangular particle}
\begin{figure}[!htbp]
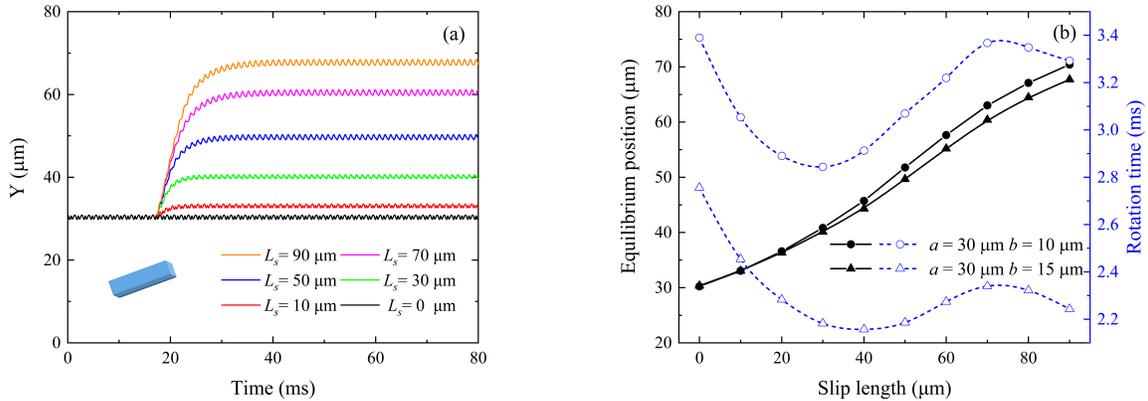

\centering
\subfigure{\includegraphics[width=8cm]{Fig_11a}}
\subfigure{\includegraphics[width=8cm]{Fig_11b}}
\caption{The migrations and equilibriums of rectangular particles in the channel flows with various unilateral slip lengths. (a) the migration trajectories and (b) the vertical equilibrium positions and the rotating periods of the rectangular particles, $a$ and $b$ are the length and width of the rectangle.}
\label{Fig. 11}
\end{figure}

The motion of rectangular particles with the size of 30 $\times$ 15 $\upmu$m and 30 $\times$ 10 $\upmu$m at different slip length are considered. FIG. \ref{Fig. 11}(a) shows that the trajectories have similar oscillation amplitudes as the elliptical particle. The migration trajectories of the rectangle ones are quite similar to the classical Segr$\rm{\acute{e}}$--Silberberg effect, although their unsymmetrical geometries cause additional fluctuations and nonuniform rotation in the process of migration. Similarly, the rectangular particle will reach different equilibrium positions with different slip lengths.

FIG. \ref{Fig. 11}(b) shows that with the increase of the slip length, the vertical equilibrium positions also gradually move towards the upper channel wall. This is in agreement with the previous studies on the lateral migrations of circular or elliptical particles. Both the elliptical and rectangular particle have the non-circular geometry, the relation between the rotation period and aspect ratio of the rectangular particle is same as that of the ellipse.
\begin{figure}[!htbp]
\centering
\includegraphics[width=100mm]{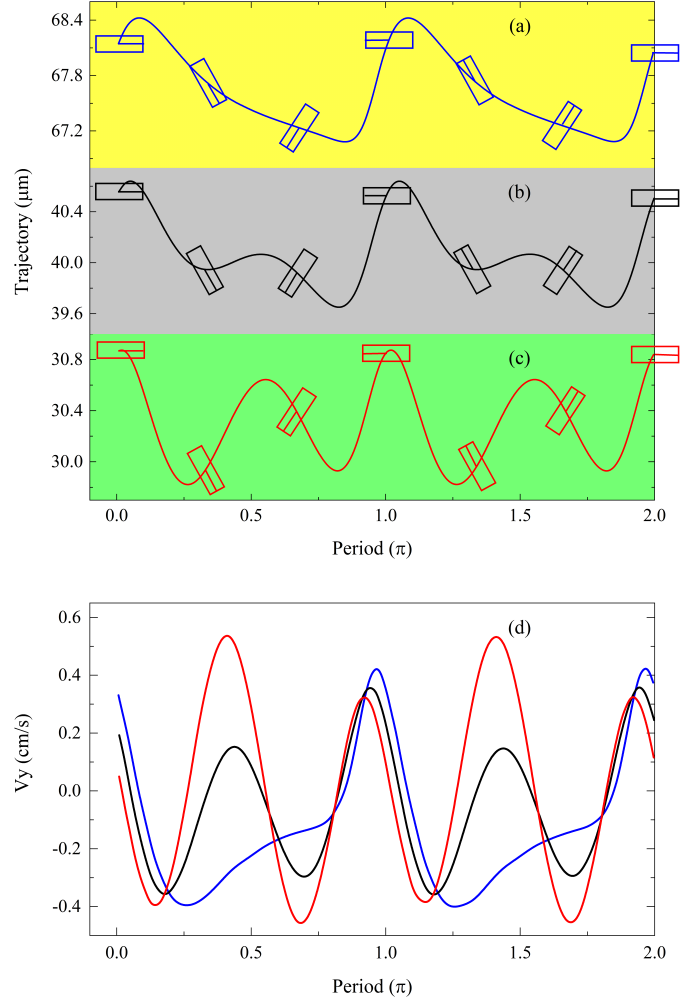}
\caption{(a) The migration trajectory and angular orientation of the rectangular particle with slip length of 90 $\upmu$m, (b) 30 $\upmu$m and (c) 0 $\upmu$m; (d) the Linear velocity of the rectangular particles with slip length of 0 (red line), 30 (black line) and 90 (blue line) $\upmu$m.}
\label{Fig. 12}
\end{figure}

FIG. \ref{Fig. 12} shows the steering angle, vertical trajectory, and linear velocity of the rectangular particles with slip length of 0, 30, 90 $\upmu$m and the 2D pose corresponding to the steering angle in a rotation period. With the increase of the slip length, the oscillation of the rectangular particle gradually changes from three wave peaks to one. When the slip length is 0 $\upmu$m and the steering angle is 0 -- 0.27$\uppi$ in half a rotation period, the alternating force caused by the rotation of the rectangular particles acts as a sinking force to make the particles move downward. However, the alternating force becomes an ascending force when the steering angle is 0.27 -- 0.54$\uppi$. In the same way, during the steering angle is 0.54 -- 0.82$\uppi$ and 0.82 -- 1$\uppi$, the alternating force makes the particle go through the ups and downs again. In the angular range of 0.27 -- 0.82$\uppi$ and 1.27 -- 1.82$\uppi$, the oscillation of the particle can be seen clearly without the slip length. When the slip length is 30 $\upmu$m, the small wave peak is not obvious. Increase to 90 $\upmu$m, the small wave peak disappears completely. These prove that the oscillation of the rectangular particle changes with the increase of the slip length.

The vertical linear velocity of the rectangular particle with different slip lengths are shown in FIG. \ref{Fig. 12}(d). Ignoring the differences in rotation periods, the patterns of linear velocities are different within a single period. It suggests that the rectangular particles have different linear motions at different slip lengths. When the steering angle is 0.27 -- 0.82$\uppi$ or 1.27 -- 1.82$\uppi$, the negative linear velocity first increases to positive and then decreases to negative when the slip length is 0 $\upmu$m. So the oscillation of rectangular particle is obvious during this period if the slip length is small. However, the boundary slip causes the acceleration of the velocity of the flow field and results in the shortening of the particle rotation period. The negative linear velocity cannot increase to positive with the increase of the slip length, leads to the fluctuation of particles is not obvious.
\begin{figure}[!htbp]
\centering
\includegraphics[width=100mm]{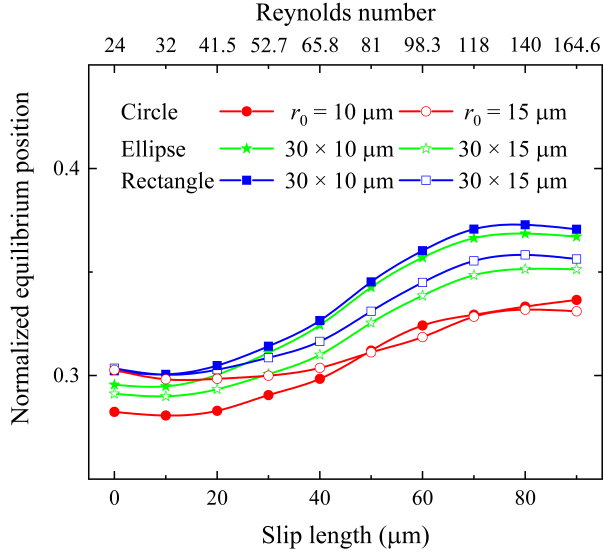}
\caption{The normalized equilibrium positions of particles in the channel widened by the relevant slip length. The red, green and blue line represents the circle, ellipse and rectangle particle, respectively. }
\label{Fig. 13}
\end{figure}

In FIG. \ref{Fig. 8}(b), FIG. \ref{Fig. 9}(b) and FIG. \ref{Fig. 11}(b), we found that the rotating periods of particles did not decrease linearly with the increasing of slip length. Considering the parabolic velocity distribution of Poiseuille flow, there are large differences of velocity among stream layers. FIG. \ref{Fig. 13} shows that with the increase of the slip length, the particles' normalized equilibrium positions gradually move from 0.3 to 0.35. That is, with the increase of Reynolds number, the normalized equilibrium position gradually moves upwards. When the slip length is less than 40 $\upmu$m, the normalized equilibrium position is maintained at about 0.3. The particle is close to the wall, the difference of the velocity between the upper and lower stream layers around the particle is large. Thus, the rotation period of particles reduces in this interval. When the slip length increases from 40 to 70 $\upmu$m, the normalized equilibrium position increase to 0.35. The particle gradually moves away from the wall, the difference of velocity between the upper and lower stream layer around the particle decreases. This result in the rotation period of the particle to be longer. So, it is clearly shown in FIG. \ref{Fig. 8}(b), FIG. \ref{Fig. 9}(b) and FIG. \ref{Fig. 11}(b) that the rotation periods of three particles increase gradually when the slip length increases from 40 to 70 $\upmu$m. 
\section{\label{sec5}Conclusion}
In this paper, we utilize the boundary slip effect to regulate the velocity distribution of fluid in the microchannel and present a scheme to active control the particle position in inertial microfluidics. A series of numerical simulations of different shape particles migrating laterally in Poiseuille flow with BSR slip boundary condition are performed by the lattice Boltzmann method with single relaxation time. The hydrodynamic force is evaluated by the Galilean-invariant momentum exchange method. The flow field of a microchannel with a unilateral slip boundary is equivalent to that of the microchannel widened by the relevant slip length, and the particle equilibrium positions in the two microchannels are consistent consequently. The effectivity and feasibility of the model are verified by three kinds of particles, namely circle, ellipse and rectangle. In addition, because the elliptical and rectangular particles have non--circular geometry, their motions are accompanied with rotation and regular vibration. The flow field and the particle motion are affected by the slip boundary conditions. When the slip boundary condition is used, the change of velocity distribution of the flow field results in accelerating the speed of particles rotation and shortening the period of rotation. It can be found that the fluctuation amplitude of the non--circular particles will increase as the slip length gradually increases. Under the slip boundary condition, some original fluctuations caused by the rotation of the particles will also disappear.

Boundary slip phenomenon has attracted increasing attention for liquid and gas flow with development of nanotechnology. The results of this work help to promote active manipulation of particles in microfluidics and realize more accurate focusing, separating and transport.

\begin{acknowledgments}
This work was supported by the National Natural Science Foundation of China (Grant Nos. 11862003, 81860635, and 12062005), the Project of Guangxi Natural Science Foundation (Grant Nos. 2017GXNSFDA198038 and 2018JJA110023), Guangxi ``Bagui Scholar'' Teams for Innovation and Research Project, Guangxi Collaborative Innovation Center of Multi-source Information Integration and Intelligent Processing.
\end{acknowledgments}


%

\end{document}